# Optomechanical measurement of photon spin angular momentum and optical torque in integrated photonic devices

Li He[1*], Huan Li[1*], and Mo Li[1†]

[1]Department of Electrical and Computer Engineering, University of Minnesota, Minneapolis, MN 55455, USA

**Photons carry linear momentum, and spin angular momentum when circularly or elliptically polarized. During light-matter interaction, transfer of linear momentum leads to optical forces, while angular momentum transfer induces optical torque[1,2]. Optical forces including radiation pressure and gradient forces have long been utilized in optical tweezers and laser cooling[3–5]. In nanophotonic devices optical forces can be significantly enhanced, leading to unprecedented optomechanical effects in both classical and quantum regimes[6–11]. In contrast, to date, the angular momentum of light and the optical torque effect remain unexplored in integrated photonics. Here, we demonstrate the measurement of the spin angular momentum of photons propagating in a birefringent waveguide and the use of optical torque to actuate rotational motion of an optomechanical device. We show that the sign and magnitude of the optical torque are determined by the photon polarization states that are synthesized on the chip. Our study reveals the mechanical effect of photon's polarization degree of freedom and demonstrates its control in integrated photonic devices. Exploiting optical torque and optomechanical interaction with photon angular momentum can lead to torsional cavity optomechanics and optomechanical photon spin-orbit**

---

[*] These authors contributed equally to this work.
[†] Corresponding author: moli@umn.edu

**coupling**[12,13], **as well as applications such as optomechanical gyroscope**[14] **and torsional magnetometry**[15]**.**

In vacuum, the linear momentum of a photon depends on its frequency $\omega$ as given by $\hbar k = \hbar \omega / c$, where $k$ is the magnitude of the wave vector and $c$ is the speed of light. When the photon is right (left) circularly polarized (defined in Fig. 1a) it also carries spin angular momentum of $+\hbar(-\hbar)$ which, noteworthily, is independent of $\omega$. The transfer of linear momentum of photons during light-matter interaction leads to optical forces including radiation pressure and gradient force. The gradient optical force can be particularly strong in nanophotonic devices which afford tight spatial confinement and strong optomechanical dispersion that can be rationally engineered[16–19]. The mechanical interaction between light and motional devices recently has been extensively explored in various optomechanical systems with the most notable achievement of backaction cooling of cavity optomechanical systems to the quantum ground states[9,20]. The other essential mechanical attribute of light, the angular momentum of photons, however, has not been investigated or utilized in integrated optomechanical systems.

Historically, the first measurement of the angular momentum of light was carried out by R. Beth in 1930s, more than three decades after the first measurement of the radiation pressure[21–24]. Beth's famous experiment was based on Poynting's prediction that circularly polarized light should exert a torque on a half-wave plate which changes the helicity of the polarization state[1], as illustrated in Fig. 1a. By detecting the rotation of a wave plate suspended on a quartz fiber, forming a torsional pendulum, as circularly polarized light passing through, Beth measured the optical torque and confirmed the angular momentum of light predicted by theory. Now, eight decades later, we are in the era of integrated photonics with lasers available as sources of coherent light. We set out to measure the spin angular momentum of light and use optical torque effect in on-chip optomechanical systems. Spin angular momentum of light is important in that it directly relates to the polarization state of light, which is an essential state variable in both classical and quantum optical information processing[25–27] In integrated photonic devices, the polarization state of light is frequently manipulated so that angular momentum exchange between the photons and the devices is ubiquitous[26–29]. Understanding and utilizing photon angular momentum induced mechanical effects should have important implication to these areas and spawn new principles of operation.

Instead of a wave plate, we consider a rectangular waveguide supporting two modes designated as transverse electric (TE) and transverse magnetic (TM) modes, resembling the two orthogonal linear polarizations in free space, as illustrated in Fig. 1b. The two modes are quasi-linearly polarized with the major components of the electric field aligned to the $x$ and $y$ axis, respectively. The magnitudes of the wave vectors of the two modes are denoted as $k_{x,y} = n_{x,y}\omega/c$, where $n_{x,y}$ is the effective mode index. In contrast to a wave plate made of anisotropic material, a rectangular waveguide of isotropic material such as silicon can have particularly strong birefringence: $\Delta n = n_x - n_y$ and $\Delta k = k_x - k_y$, due to its geometric anisotropy. Therefore, waveguides are frequently engineered to manipulate the polarization state on a chip[28]. We use the Jones vector to denote the polarization state of an arbitrary hybrid mode as: $\begin{pmatrix} a_x & a_y e^{-i\varphi} \end{pmatrix}^T$, where $a_{x,y}$ and $\varphi$, respectively, are the mode amplitude and phase difference of the electric field components of the TE and TM modes so that $a_{x,y}^2 = P_{TE,TM}$ is the power of each mode. The polarization state is linear when $\varphi = 0$ or $\pi$; right (left) circular when $a_x = a_y$ and $\varphi = -(+)\pi/2$. Because of birefringence, $\varphi$ is a function of propagation distance $\varphi(z) = \varphi_0 + \Delta k z$ so that the polarization state of light, if not purely TE or TM ($a_x a_y \neq 0$), keeps evolving as light propagates along the waveguide (Fig. 1b). Like the wave plate situation, accompanying the evolution between linear, elliptical and circular polarization states is angular momentum exchange between the light and the waveguide. As a result, the photons apply an optical torque on the birefringent waveguide to twist it. The total optical torque $T(l)$ on the waveguide of length $l$ should be the integration of the time-averaged, local linear torque density $\tau(z)$, which varies along the waveguide as the polarization state evolves and can be express as: $\tau(z) = -\Phi_q \left(dS_e/dz\right)$, where $\Phi_q$ is photon flux and $S_e$ is the effective spin angular momentum of a photon. Such a principle is illustrated in Fig. 1b and c. In more detail, when light of arbitrary polarization state propagates along the waveguide, the electromagnetic field generates a distributed force, hence a distributed torque $\tau_z(x, y, z)$, inside the waveguide. A representative torque distribution inside a typical waveguide cross-section where $\varphi = 0$ is shown in Fig. 1d. The integration of this distributed torque over the waveguide cross-section yields torque per unit length:

$$\tau(z) = -\Phi_q \frac{dS_e}{dz} = \eta \frac{\Delta k}{\omega} \left(2 a_x a_y\right) \cos(\varphi(z)) \ , \tag{1}$$

which leads to an expression of $S_e$:

$$S_e(\varphi) = -\eta \hbar \frac{2 a_x a_y}{a_x^2 + a_y^2} \sin(\varphi) \ . \tag{2}$$

The coefficient $\eta$ equals 1 in vacuum and in dielectric materials, it accounts for interactions including dipole forces[2] and electrostrictive forces[30]. From Eq. (2), it is evident that $S_e = \pm \eta \hbar$ when the polarization state is circular, while $S_e = 0$ when the polarization state is linear. In the Supplementary Information (SI), we have included a detailed theoretical analysis of the waveguide case, which shows that, for a waveguide made of silicon, the dipole and electrostrictive contributions to the optical torque are on the same order of magnitude and dependent on the waveguide geometry and material properties.

## Results

**Torsional cavity optomechanical resonator**

To demonstrate and measure the photon spin angular momentum and optical torque, we fabricated silicon nano-optomechanical devices using silicon-on-insulator (SOI) wafers. Figure 2a shows an optical image of the device. The core element of the device is a two-mode waveguide suspended from the substrate. With a rectangular cross-section of 400 nm in width (along the *x* direction) and 340 nm in height (along the *y* direction), the waveguide was designed to have a strong birefringence for its TE and TM modes with $\Delta n = 0.16$. Attached to the waveguide and also suspended from the substrate is a nanobeam in which two one-dimensional photonic crystal nanocavities are embedded, one on each side. The resonance modes of the nanocavities are optimized to be particularly sensitive to the distance to the substrate so they provide very sensitive detection to the rotation of the waveguide actuated by the optical torque[31]. Therefore, the nanobeam with nanocavities is analogous to the mirror mounted on the quartz fiber in Beth's apparatus. One of the nanocavities is coupled to a nearby waveguide (inset, Fig. 2c). The representative transmission spectrum through this coupling waveguide shown in Fig. 2b exhibits the resonance of the nanocavity, with a loaded (intrinsic) quality factor of 4,200 (23,000).

**Controlling optical torque through synthesis of polarization state**

The optical torque on the waveguide can be controlled by varying the polarization parameters $(a_x, a_y, \varphi)$. This was achieved with on-chip peripheral photonic circuits and an off-chip setup, as illustrated in Fig. 2c. Specifically, the TE and TM modes were derived from the same laser source and their power ratio was controlled by a fiber polarization controller (FPC1). After the electro-optic phase modulator (EOM), the TE and TM modes were separated by a fiber polarization beam splitter (PBS) into two optical paths, where their polarization states were independently conditioned and their power monitored. In order to accurately control and stabilize the relative phase $\varphi$, an on-chip integrated interferometer provided the feedback, which was particularly effective for real-time compensation of the phase fluctuations in the optical fibers caused by room temperature fluctuations. The output power of the interferometer ($P_I$) is determined by $\cos(\varphi_I)$, where $\varphi_I$ is the relative phase between the two interferometer arms. Although the actual difference between $\varphi_I$ and $\varphi$ cannot be practically predetermined by design, the difference $\varphi - \varphi_I$ remains consistent for any single device at a stabilized temperature. Therefore, $P_I$ was measured and used to control the bias of the EOM and a fiber thermo-optic phase shifter in the TM optical path, compensating, respectively, fast and slow fluctuations of the optical phases in the system. At the same time, the measurement system was in an enclosure with temperature fluctuations controlled to be within ±0.2 K.

**Optical torque modulation and measurement**

To measure the optical torque, we employed the resonance method used in Beth's original experiment but in a modern fashion, by modulating the optical torque to actuate resonant motion of the nanobeam hinging on the waveguide (Fig. 3a). To do so, we utilized the different modulation efficiency of the EOM for the TE and TM modes, which leads to relative phase modulation between the two modes. To understand the mechanical motion of the waveguide-nanobeam structure actuated by the optical torque, Fig. 3b shows the simulated fundamental torsional mode of the structure, which is the dominant mode that is excited by the optical torque. In this mode, the waveguide undergoes pure torsional motion with its two ends fixed while the nanobeam tilts with it. This mode can be described as an effective torsional simple harmonic oscillator, the effective driving torque of which, $T_e$, is defined by the overlap integral between the optical linear torque density $\tau(z)$, given in equation (1), and the normalized angular mode profile $\theta_n(z)$ along the waveguide as: $T_e = \int_{-l/2}^{l/2} \tau(z)\theta_n(z)dz = T_m \cos(\varphi_0)$, where $\varphi_0 = \varphi(z=0)$ and the integration is simplified because $\theta_n(z)$ is an even function of $z$. Therefore, the effective torque $T_e$ changes sinusoidally with $\varphi_0$ and reaches extrema when $\varphi_0$ equals to 0 or $\pi$. The maximal effective torque $T_m$ depends on the waveguide length and birefringence, and after approximating $\theta_n(z)$ to a triangular function (see SI) can be expressed as:

$$T_m \approx -S_e(\frac{\pi}{2})\Phi_q \frac{2\sin^2(\Delta kl/4)}{\Delta kl/4} . \qquad (3)$$

To confirm above theory and achieve quantitative measurement of optical torque, we measured devices with various suspended waveguide length $l$. Fig. 3c shows the noise power spectral density (PSD) measured from the transmitted probe laser power of a representative device with $l = 10.5$ μm. Four mechanical resonance peaks due to thermo-mechanical fluctuation are observed with the peak at 358.7 kHz corresponding to the fundamental torsional mode, which is plotted in detail in Fig. 3d, showing a quality factor of 12,000. We used such thermo-mechanical noise PSD measurement to calibrate our system's measurement sensivity[32]. The measured resonance frequencies of seven devices with various waveguide length $l$ = 9.0, 10.5, 12.0, 13.5, 15.0 16.5 and 18.0 μm are summarized in Fig. 3e. The results agree well with the simulation results (blue line) using an elasticity matrix for silicon 19% lower than the typical value for bulk silicon (red line)[33].

When performing resonance measurement, the EOM was driven by a network analyzer to modulate the phase $\varphi_0$ sinusoidally by a small amplitude of $\delta\varphi_0$ and thus generate a dynamic effective torque on the waveguide: $\delta T_e \cos(\Omega t) = -T_m \sin(\varphi_0) \delta\varphi_0 \cos(\Omega t)$. Importantly, by performing only phase modulation and avoiding amplitude modulation, the photothermal effect was minimized[8,34,35]. Additionally, the structural symmetry of the device and the large gap between the waveguide and the substrate eliminated the "light pressure torque" effect[21], even though a small residual amplitude modulation might remain due to instrumental non-ideality. We first measured the nanobeam's resonance responses with constant $\delta\varphi_0$ and optical power $P$ (and $a_x = a_y$) while $\varphi_0$ was varied (through controlling $\varphi_1$). The quadrature components of the responses from a representative device ($l$ = 10.5 µm) are plotted in Fig. 4a, showing that the fundamental torsional mode of the nanobeam is excited by the optical torque. The data measured for six different values of $\varphi_0$ clearly reveal that both the amplitude and sign of $\delta T_e$ depend on $\varphi_0$. In the experiment, the actuated mechanical rotation amplitude was kept low at a few µrad to avoid inducing nonlinearity. From the response for each value of $\varphi_0$, $\delta T_e / \delta\varphi_0$ is calculated and plotted in Fig. 4b, showing a perfect sinusoidal dependence predicted by $-T_m \sin(\varphi_0)$, from which the value of $T_m$ is obtained.

To further investigate the dependence of $T_m$ on the polarization state, we first varied the ratio between $P_{TE} = a_x^2$ and $P_{TM} = a_y^2$ with constant total power $P = a_x^2 + a_y^2 = 95\ \mu W$. From the measurement of two devices ($l$ = 10.5 and 9 µm), we observed a semi-circular dependence shown in Fig. 4c, which originates from the $(2a_x a_y)/(a_x^2 + a_y^2)$ term in equation (2). With the same two devices, we also measured the dependence of $T_m$ on power $P$ with $a_x = a_y$, shown in Fig. 4d. The results show that the optical torque $T_m$ deviates from linear dependence on power at above 100 µW. We attribute the behavior to mechanical nonlinearity of the device, which requires further investigation but does not affect the measurement results obtained within the linear region. Finally, we investigated the dependence of $T_m$ on waveguide length $l$ with fixed power (95 µW and $a_x = a_y$). The result in Fig. 4e is fitted to equation (2) and (3) with $\eta$ and $\Delta n$ as the free parameters, showing a good agreement with the theory. From the fitting, the $\Delta n$ obtained is 0.18, very close to the designed value of 0.16, while the $\eta$ obtained is 2.2±1.0. Therefore, the measured value of photon spin angular momentum in the waveguide is $(2.2 \pm 1.0)\hbar$. The theoretical value of $\eta$ is calculated to be 1.5 (see SI), which thus is within the measurement uncertainty of the experiment.

## Discussion

Our experiment provides the first unambiguous measurement of spin angular momentum of photons and optical torque generated in an integrated photonic device. The result shows that the optical torque is determined by the geometric birefringence of the waveguide, which can be rationally engineered to enhance the effect. Such designs can be achieved in other nanophotonic structures such as plasmonic devices, metasurfaces and metamaterials to generate even more pronounced effects[36]. Further, since the photon angular momentum is only dependent on the polarization state and independent of the optical frequency, the optical torque effect is universal over a broad spectral band, providing a large leeway for engineering. Our experiment measured the photon angular momentum inside a waveguide, and the fact that the value we measured is more than the vacuum value suggests that in our system, the Minkowski photon angular momentum[37] is applicable and the electrostriction effect[30] does contribute to the photon spin angular momentum. In addition to spin angular momentum, photons can also have orbital angular momentum and optomechanical effects arising from the spin-orbit interactions of light in nanophotonic systems will be even more intriguing[12].

**Figures**

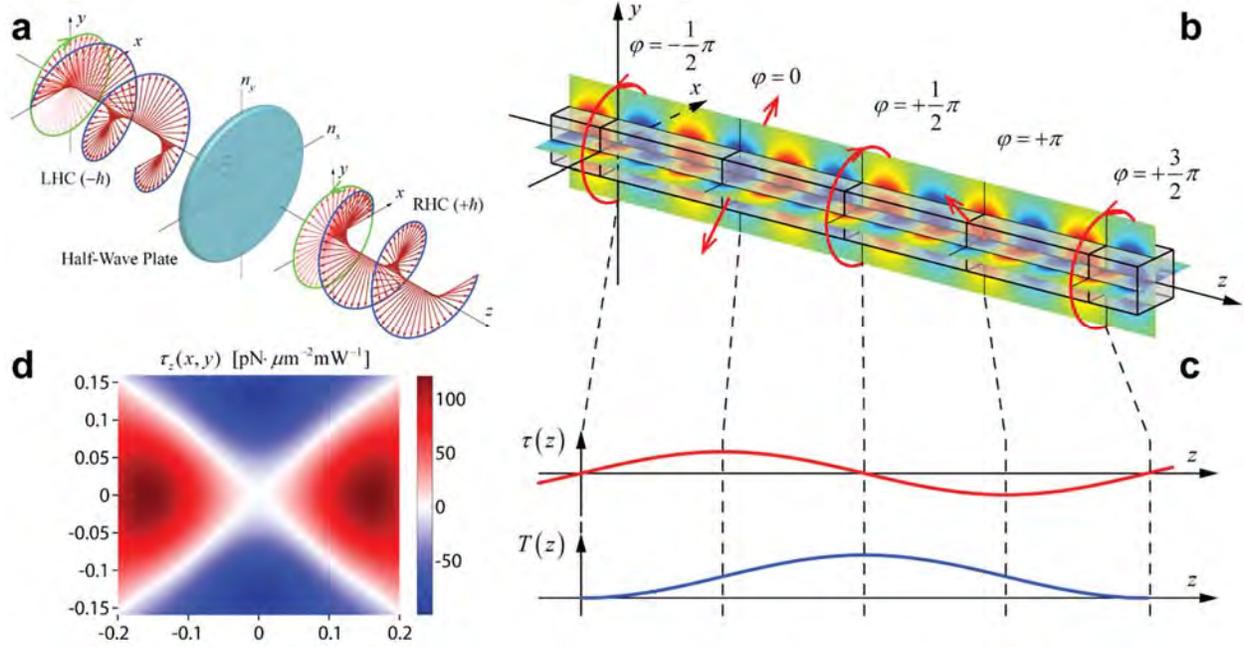

**Figure 1 Polarization conversion and optical torque in a wave plate and a birefringent waveguide. (a)** After passing through a half-wave plate, right-handed circularly (RHC) polarized light is converted to left-handed circularly polarized (LHC) and transfers angular momentum of $2\hbar$ per photon to the wave plate and rotates it. Beth's experiment measured photon angular momentum by detecting the rotation of the wave plate suspended as a torsional pendulum. **(b)** The polarization state of a hybrid mode in a rectangular waveguide evolves continuously from circular to elliptical to linear as it propagates along the waveguide. The horizontal (vertical) plane plots the $E_x$ ($E_y$) component of TE (TM) mode. **(c)** Linear torque density distribution $\tau(z)$ and total torque $T(z)$, which is the integration of $\tau(z)$, vary sinusoidally along the waveguide. **(d)** Distribution of the $z$-component of the torque density $\tau_z(x,y)$ on the waveguide cross-section applied by the optical mode including contributions from dipole and electrostrictive forces. The waveguide geometry is assumed to be 400 nm wide and 340 nm high and the phase $\varphi=0$.

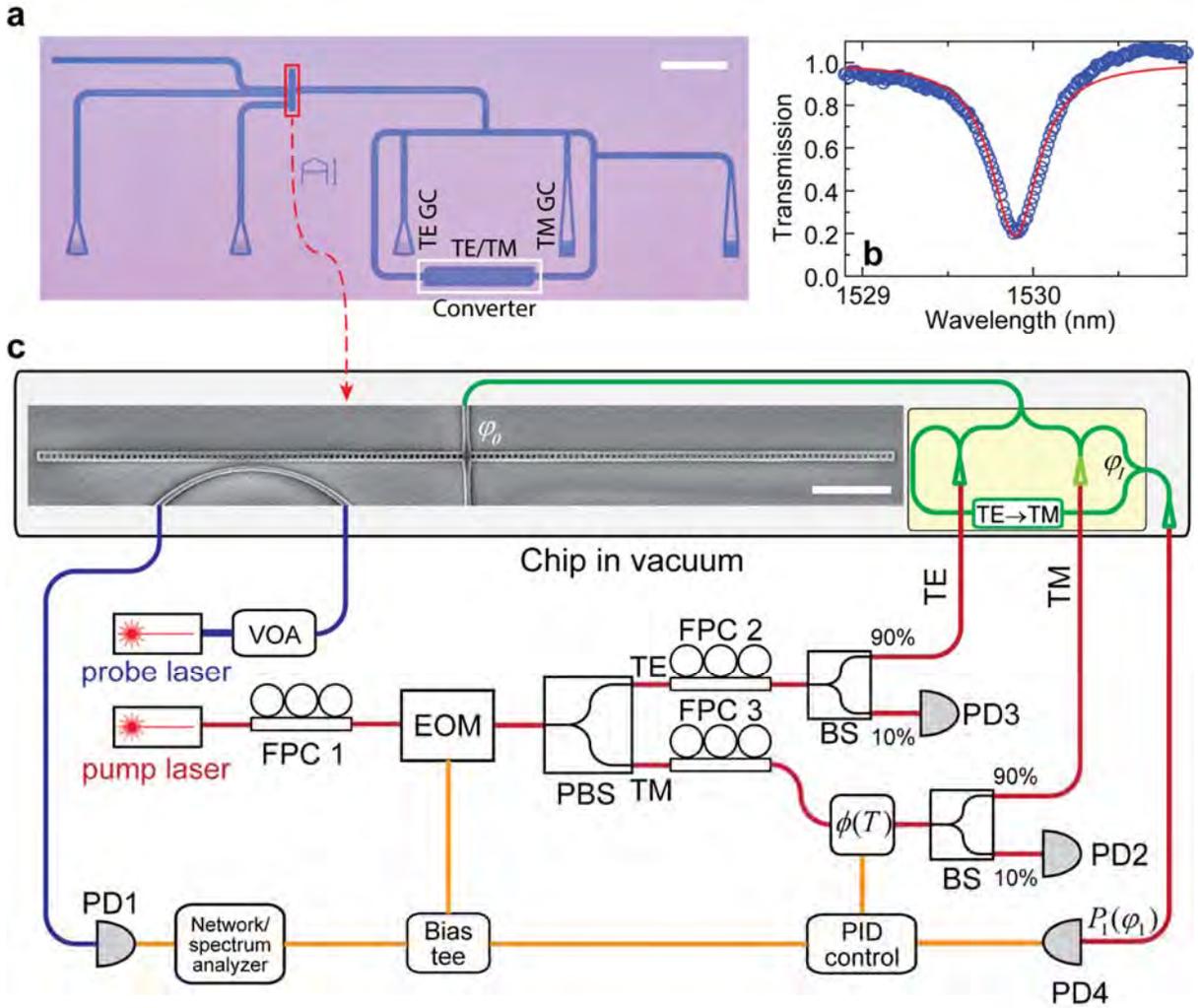

**Figure 2 Optomechanical scheme to measure photon angular momentum and optical torque in a waveguide**. (a) Optical microscope image of the complete integrated photonic device including the suspended waveguide-nanobeam structure (in the red box), TE and TM mode grating couplers (GC) and an interferometer with a TE to TM mode converter (inside the white box) for phase stabilization. Scale bar: 100 μm. (b) Optical transmission spectrum measured from the waveguide coupled to one of the nanocavities on the nanobeam. The resonance mode having a loaded (intrinsic) quality factor of $4.2\times10^3$ ($2.3\times10^4$) is optimized for detecting the rotation of the nanobeam. (c) Schematic of the measurement system. FPC: fiber polarization controller; EOM: electro-optic phase modulator; BS: beam splitter; PBS: polarization beam splitter; PD: photodetector; VOA: variable optical attenuator; $\phi(T)$: fiber thermo-optic phase shifter. Inset: scanning electron microscope image of the waveguide and the nanobeam with photonic crystal cavities. Scale bar: 5 μm.

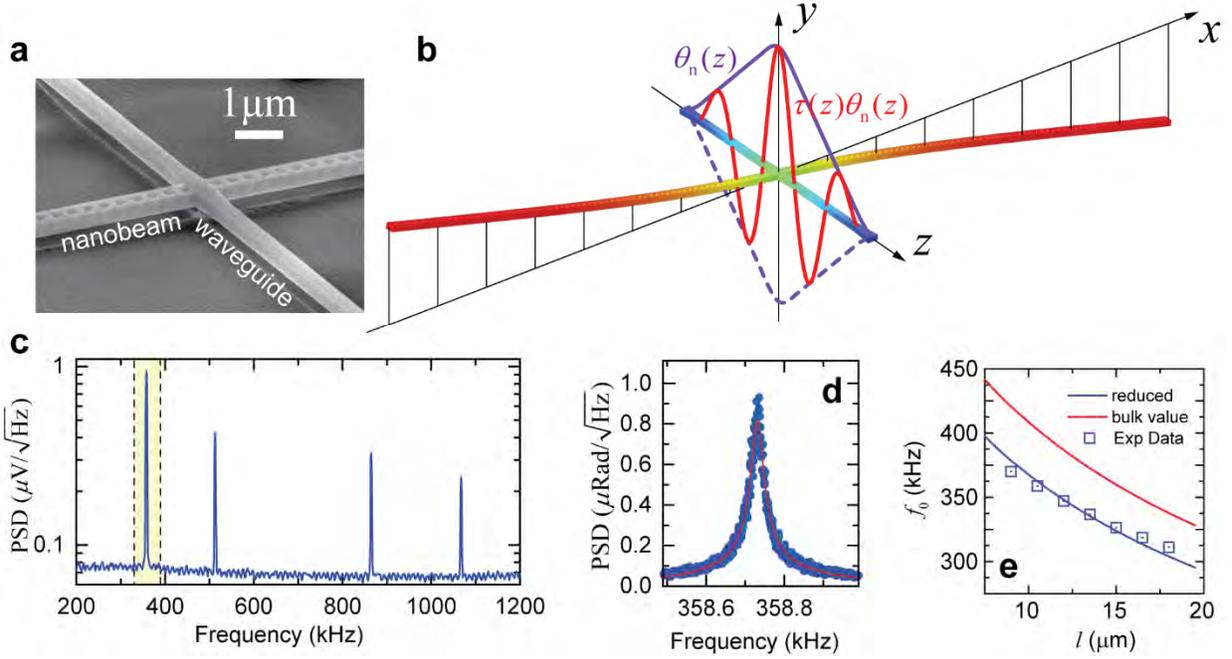

**Figure 3 Mechanical characteristics of the torsional optomechanical device. (a)** Zoom-in scanning electron microscope image showing the suspended waveguide-nanobeam structure. **(b)** Simulated fundamental torsional mode of the waveguide-nanobeam structure. Also plotted on the *y-z* plane are the normalized angular mode profile $\theta_n(z)$ and its representative overlap integrand with torque distribution: $\theta_n(z) \cdot \tau(z)$, integration of which along the waveguide yields effective torque $T_e$. **(c)** Broadband thermo-mechanical noise power spectral density (PSD) measured with the probe nanocavity resonance, showing four prominent mechanical modes. They are, with increasing frequency, fundamental (or first-order) torsional, fundamental (or first-order) flapping, second-order torsional and second-order flapping modes. **(d)** Zoom-in view of the fundamental torsional mode resonance at 358.7 kHz, showing a quality factor of 12,000. The PSD unit has been calibrated and converted to rotational angle. **(e)** Fundamental torsional resonance frequency of devices with various waveguide length *l*. The measurement results show lower frequency than simulation using typical bulk value of silicon's elasticity matrix (red line), indicating the elasticity matrix of the silicon layer in the SOI is effectively lower (blue line).

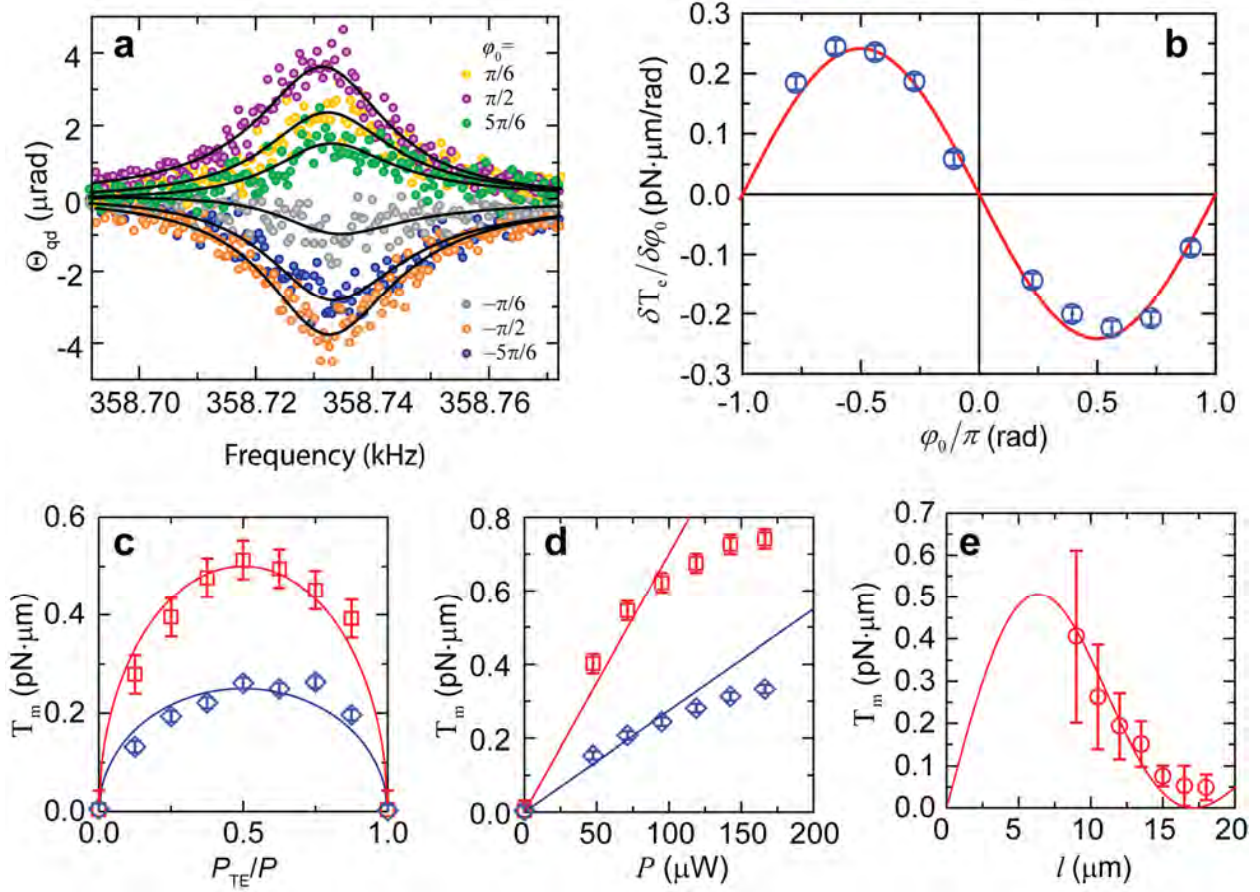

**Figure 4 Measurement of angular momentum and optical torque applied on the waveguide by light of arbitrary polarization state.** **(a)** The quadrature resonance responses of the device with $l$ = 10.5 μm at its fundamental torsional mode when the phase $\varphi_0$ is varied and the optical power is kept at 95 μW. The results show both magnitude and sign of the driving optical torque is changed with $\varphi_0$. **(b)** The measured amplitude of modulated torque $\delta T_e$ over phase modulation $\delta\varphi$ (blue symbols) depends sinusoidally (red line) with the phase $\varphi_0$ as in equation (2). The maximum torque $T_m(l)$ is determined from this result. **(c)** Optical torque $T_m(l)$ measured (symbols) when the fraction of TE mode power $P_{TE}$ is varied while the total power $P$ is kept constant, for two devices with $l$ = 10.5 (blue) and 9 (red) μm. The results show predicted semi-circular dependence (lines). **(d)** Optical torque $T_m(l)$ measured (symbols) when the total power $P$ is varied, for the same two devices. The results show linear dependence below 100 μW (lines). **(e)** Optical torque $T_m(l)$ for devices with various waveguide length $l$. Fitting the data with theoretical

model (line) yields the waveguide birefringence $\Delta n$ and the coefficient of angular momentum transfer per photon $\eta$.

## Method

**Device fabrication:** The devices were fabricated on a SOI wafer with a 340 nm thick device layer and a 3 μm buried oxide layer. The silicon structures were first patterned using electron beam lithography (Vistec EBPG5000+) and fluorine-based plasma dry etching processes. The electron beam resist ZEP520A was developed at −15 °C after the exposure to enhance the resolution. The critical dimensions of the silicon structure were confirmed with SEM to be within ±5 nm of the designed values. The waveguide and nanobeam were subsequently released from the substrate in two steps using photolithography and wet etching processes. In the first step, only the substrate below the waveguide was selectively etched by 210 nm with buffered oxide etch (BOE), while in the second step, the substrate below both the waveguide and the nanobeam was etched by 330 nm with diluted hydrofluoric (HF) acid. The resultant gap below the waveguide is 540 nm, large enough to decouple the waveguide modes from the substrate and to minimize gradient optical force, while the gap below the nanobeam is 330 nm, small enough to enable sensitive transduction of the torsional motion. After wet etching, we dried the sample in a critical point dryer to avoid the stiction problem during solvent evaporation.

**Optical torque measurement:** To minimize air damping, the devices were measured in a vacuum chamber with pressure less than $1 \times 10^{-4}$ Torr. A fiber array was used to access them through the on-chip grating couplers. All the off-chip experimental equipment that is sensitive to the temperature fluctuations was enclosed in a plastic shield. In the experiment, the temperature fluctuations inside the plastic shield and the vacuum chamber were controlled to be within ±0.2 K.

The probe laser (blue path in Fig. 2c) was sufficiently attenuated to avoid dynamic cavity optomechanical backaction and optimally detuned from the nanocavity resonance to transduce the torsional motion. The width of the coupling waveguide for the nanocavity is tapered down to 320 nm to achieve phase matching with the cavity mode. The probe laser output was measured by PD1. The pump TE and TM modes were derived from the same laser source (red path in Fig. 2c) for optimal coherence between them. The FPC1 was used to tune the power ratio of the TE and TM modes in the EOM (Lucent 2623NA) coupled from the linearly polarized laser source. The half-wave voltage of the EOM ($V_\pi$) is 3 V for the TE mode and 9 V for the TM mode. Therefore, the voltage applied on the EOM linearly modulates the phases of both modes with different efficiencies, equivalently modulating $\varphi_0$ with $V_\pi = 4.5$ V. The fiber PBS was used to separate the TE and TM modes of the EOM into two optical paths, where their polarization states were conditioned with FPCs for their respective grating couplers and their power was tapped (10%) with beam splitters and monitored with photodetectors, before they were coupled on-chip. With the beam splitters and combiners in the on-chip peripheral photonic circuits (green path in Fig. 2c), half of the TE and TM mode power were split from their respective input grating couplers and combined into the suspended waveguide, where the induced optical torque was measured with the nanobeam. The junction between the suspended waveguide and the nanobeam is optimized for minimal perturbation to the propagation of both modes inside the waveguide. After the suspended section, the waveguide width is adiabatically tapered down to 80 nm so that the pump laser was emitted into free space with minimal reflection. To implement the on-chip integrated interferometer that provides the feedback used to control and stabilize $\varphi_0$, the other half of the TE and TM mode power were guided into their respective interferometer arms, where the TE mode was converted into pure TM mode with an on-chip mode converter and filter. The interferometer output $P_I$ is determined by $\varphi_I$ and the optical power in the two interferometer arms, $P_{I1}$ and $P_{I2}$, as

$$P_I = \frac{P_{I1}}{2} + \frac{P_{I2}}{2} + \sqrt{P_{I1}P_{I2}}\cos(\varphi_I),$$

which was measured by PD4. The extinction ratio of the interferometer, defined as the ratio between the maximum and minimum output power $P_{max}/P_{min}$, is determined by $P_{I1}/P_{I2}$ as

$$ER = \frac{P_{max}}{P_{min}} = \left(\frac{\sqrt{P_{I1}/P_{I2}}+1}{\sqrt{P_{I1}/P_{I2}}-1}\right)^2.$$

In practice, our feedback and control scheme performed satisfactorily when $ER > 1.5$, requiring that $0.01 < P_{I1}/P_{I2} < 100$, which was naturally met in most of the situations. A minor limitation of this feedback scheme, however, is that it fails to function when $|\cos(\varphi_I)| \to 1$, where the dependence of $P_I$ on $\varphi_I$ becomes insensitive and non-monotonic. In the actual experiment, we always kept $|\cos(\varphi_I)| \leq \sqrt{3}/2$, which provided satisfactory phase stabilization and was more than sufficient to reveal the sinusoidal dependence shown in Fig. 4b.

Electronic equipment (orange path in Fig. 2c) was used for both the measurement and control. A bias tee was used to combine the RF signal from the network analyzer and the bias voltage from the PID controller. We used a network/spectrum analyzer (Agilent 4396B) for the measurement. For the thermo-mechanical noise measurement shown in Fig. 3c and d, spectrum mode was used to measure the PSD of PD1 output, which was used to calibrate the optomechanical measurement transduction factors by theoretical curve-fitting. For the resonance response measurement shown in Fig. 4, network mode was used to drive the EOM through the bias tee RF port, and simultaneously measure the response signal from PD1. Meanwhile, the proportional-integral-derivative (PID) control scheme used the feedback from PD4 output to control and stabilize $\varphi_0$ by compensating the random phase fluctuations in the optical fibers in real-time. We employed a two-stage PID control scheme tailored for our experiment. The first stage consists of an analog PID controller (SRS SIM960), which directly stabilizes the PD4 output by tuning the EOM bias voltage through the bias tee DC port, whose bandwidth is more than sufficient to compensate the fastest phase fluctuations in our experiment. The second stage consists of a house-made PID controller, which minimizes the EOM bias voltage to reduce Joule heating in the EOM by tuning the fiber thermo-optic phase shifter in the TM optical path. In this configuration, the first stage provides fast response and the second stage slowly takes over any accumulated burden on the first PID stage with its large tuning range.

**Acknowledgement**

We acknowledge the funding support provided by the Young Investigator Program (YIP) of AFOSR (Award No. FA9550-12-1-0338). Parts of this work were carried out in the University of Minnesota Nanofabrication Center which receives partial support from NSF through NNIN program, and the Characterization Facility which is a member of the NSF-funded Materials Research Facilities Network via the MRSEC program. H. Li acknowledges the support of Doctoral Dissertation Fellowship provided by the Graduate School of the University of Minnesota.

**Author Contributions**

M. L. conceived and supervised the research; all authors designed the experiments; L. H. designed and fabricated the devices and performed simulation; H. L. performed the measurement; H. L. and L. H. analyzed the data; all authors wrote the paper.

**Competing Financial Interests**

The authors declare no competing financial interests.

# Optomechanical measurement of photon spin angular momentum and optical torque in integrated photonic devices

Li He[1*], Huan Li[1*] and Mo Li[1†]

[1]Department of Electrical and Computer Engineering, University of Minnesota, Minneapolis, MN 55455, USA

## Supplementary Information

### 1. Theory of linear optical torque density in silicon waveguide

Guided by the intuition that in dielectric media each photon carries spin angular momentum (on the order of $\hbar$) and the change of which leads to optical torque, the linear optical torque density can be written as,

$$\langle \tau(z) \rangle = -\Phi_q \frac{dS_e}{dz}, \tag{S1}$$

where $\langle \cdots \rangle$ denotes time averaging, $\Phi_q = P/\hbar\omega = (a_x^2 + a_y^2)/\hbar\omega$ is the photon flux and $S_e$ is the effective spin angular momentum per photon. As will be shown later, in the case of two orthogonal modes co-propagating in a silicon waveguide, the amplitude of linear optical torque density is locked to light polarization. The correlation between optical torque and light polarization manifests the transfer of photon spin angular momentum to or from dielectric media.

The force exerted by electromagnetic fields on dielectric media during light-matter interaction includes dipole and electrostrictive contributions. Similarly, in the calculation of $\langle \tau(z) \rangle$ and $S_e$, the two distinct mechanisms must be evaluated separately as we approached below,

$$\langle \tau(z) \rangle = \langle \tau^d(z) \rangle + \langle \tau^{es}(z) \rangle = -\Phi_q \left( \frac{dS_e^d}{dz} + \frac{dS_e^{es}}{dz} \right). \tag{S2}$$

### 1.a. Dipole contribution

The optical torque density that arises from dipole force on linear dielectric media can be written in the general form as[1]

$$\tau^d(\mathbf{r}) = \mathbf{r} \times \left[ (\mathbf{P} \cdot \nabla)\mathbf{E} + \frac{d\mathbf{P}}{dt} \times \mathbf{B} \right] + \mathbf{P} \times \mathbf{E}. \tag{S3}$$

---

[*] These authors contributed equally to this work.
[†] Corresponding author: moli@umn.edu

For isotropic media such as silicon, the contribution from the internal torque $\mathbf{P} \times \mathbf{E}$ vanishes, as $\mathbf{E}$ and $\mathbf{P}$ are always in the same direction. Furthermore, assuming optical modes with time dependence $e^{-i\omega t}$, the time-averaged force associated with the magnetic fields $\frac{d\mathbf{P}}{dt} \times \mathbf{B}$ also vanishes. The only non-vanishing term $(\mathbf{P} \cdot \nabla)\mathbf{E}$ needs to be treated with caution at dielectric interfaces where normal components of $\mathbf{E}$ and $\mathbf{P}$ are discontinuous. Therefore, we evaluate the torque in Eq. (S3) inside the material and at the dielectric interfaces separately as follows.

The bulk contribution is obtained by integrating Eq. (S3) over the volume inside the dielectric media

$$\mathbf{T}^{d,bulk} = \int \mathbf{r} \times (\mathbf{P} \cdot \nabla)\mathbf{E} d^3x = \frac{1}{2}\varepsilon_0(\varepsilon-1)\int \mathbf{r} \times \nabla(\mathbf{E} \cdot \mathbf{E})d^3x, \tag{S4}$$

where we employed following identity and dropped the term involving magnetic fields.

$$(\mathbf{E} \cdot \nabla)\mathbf{E} = \frac{1}{2}\nabla(\mathbf{E} \cdot \mathbf{E}) - \mathbf{E} \times (\nabla \times \mathbf{E}) = \frac{1}{2}\nabla(\mathbf{E} \cdot \mathbf{E}) + \mathbf{E} \times \frac{d\mathbf{B}}{dt} = \frac{1}{2}\nabla(\mathbf{E} \cdot \mathbf{E}). \tag{S5}$$

Therefore, the time-averaged torque along waveguide direction ($z$ direction) is

$$\langle T^{d,bulk} \rangle = \frac{1}{4}\varepsilon_0(\varepsilon-1)\int \left[ x\frac{\partial(\mathbf{E} \cdot \mathbf{E}^*)}{\partial y} - y\frac{\partial(\mathbf{E} \cdot \mathbf{E}^*)}{\partial x} \right] d^3x. \tag{S6}$$

Consider two lowest waveguide modes with quasi-TE and TM polarization are co-propagating along $z$ direction. The complex electric fields in the waveguide can be written as the sum of the two modes

$$\mathbf{E} = \mathbf{E}_{TE}(x,y)e^{ik_xz} + \mathbf{E}_{TM}(x,y)e^{ik_yz - i\varphi_0}, \tag{S7}$$

where $k_x$ and $k_y$ are the wave vectors of each mode and $\varphi_0$ is the relative phase at $z=0$. With Eq. (S7) it can be easily verified that the $\mathbf{E} \cdot \mathbf{E}^*$ term in Eq. (S6) includes four parts

$$\mathbf{E} \cdot \mathbf{E}^* = \mathbf{E}_{TE} \cdot \mathbf{E}_{TE}^* + \mathbf{E}_{TM} \cdot \mathbf{E}_{TM}^* + \mathbf{E}_{TE} \cdot \mathbf{E}_{TM}^* e^{i\varphi} + \mathbf{E}_{TM} \cdot \mathbf{E}_{TE}^* e^{-i\varphi}, \tag{S8}$$

where $\Delta k = k_x - k_y$ and $\varphi(z) = \Delta k z + \varphi_0$. We note that the first two terms in Eq. (S8) describe the contributions due to TE and TM mode alone, while the cross terms are non-vanishing only when the two modes are both present. Due to the symmetry of the field distribution, the net torque induced by TE and TM mode alone vanishes and only the cross terms contribute. Thus Eq. (S6) can be further written as

$$\langle T^{d,bulk} \rangle = \frac{1}{4}\varepsilon_0(\varepsilon-1)\int \left[ x\frac{\partial(\mathbf{E}_{TE} \cdot \mathbf{E}_{TM}^*)}{\partial y} - y\frac{\partial(\mathbf{E}_{TE} \cdot \mathbf{E}_{TM}^*)}{\partial x} \right] dxdy \int e^{i\varphi} dz + c.c.. \tag{S9}$$

We can always set the phase of the two eigenmodes in such a way that the two transverse electric field components $(E_x, E_y)_{TE,TM}$ in Eq. (S7) are real while the longitudinal component is imaginary. Therefore, Eq. (S9) can be written as

Preprint-Mo Li Group$$\langle T^{d,bulk} \rangle = \int dz \left\{ \frac{1}{2}\varepsilon_0(\varepsilon-1)\cos(\varphi) \int \left[ x\frac{\partial(\mathbf{E}_{TE} \cdot \mathbf{E}^*_{TM})}{\partial y} - y\frac{\partial(\mathbf{E}_{TE} \cdot \mathbf{E}^*_{TM})}{\partial x} \right] dxdy \right\} = \int dz \langle \tau^{d,bulk}(z) \rangle, \quad (S10)$$

where the linear torque density (torque per unit length) $\langle \tau^{d,bulk}(z) \rangle$ can be obtained as

$$\langle \tau^{d,bulk}(z) \rangle = \frac{1}{2}\varepsilon_0(\varepsilon-1)\cos(\varphi(z)) \int \left[ x\frac{\partial(\mathbf{E}_{TE} \cdot \mathbf{E}^*_{TM})}{\partial y} - y\frac{\partial(\mathbf{E}_{TE} \cdot \mathbf{E}^*_{TM})}{\partial x} \right] dxdy. \quad (S11)$$

If we normalized the field amplitudes to the mode powers such that: $\mathbf{e}_{TE,TM} = \mathbf{E}_{TE,TM}/\sqrt{P_{TE,TM}} = \mathbf{E}_{TE,TM}/a_{x,y}$, then Eq. (S11) is revised to be

$$\langle \tau^{d,bulk}(z) \rangle = \frac{1}{2}\varepsilon_0(\varepsilon-1)\cos(\varphi(z))(a_x a_y) \int \left[ x\frac{\partial(\mathbf{e}_{TE} \cdot \mathbf{e}^*_{TM})}{\partial y} - y\frac{\partial(\mathbf{e}_{TE} \cdot \mathbf{e}^*_{TM})}{\partial x} \right] dxdy. \quad (S12)$$

Remarkably, Eq. (S12) predicts that $\langle \tau^{d,bulk}(z) \rangle$ exhibits a sinusoidal dependence on $\varphi$, which is closely related to light polarization. Comparison with Eq. (S2) shows that the effective angular momentum per photon due to dipole force inside media can be written as

$$S_e^{d,bulk} = -\eta^{d,bulk} \hbar \frac{2a_x a_y}{a_x^2 + a_y^2} \sin(\varphi), \quad (S13)$$

where $\eta^{d,bulk}$ is a dimensionless coefficient:

$$\eta^{d,bulk} = \varepsilon_0(\varepsilon-1)\frac{\omega}{4\Delta k} \int \left[ x\frac{\partial(\mathbf{e}_{TE} \cdot \mathbf{e}^*_{TM})}{\partial y} - y\frac{\partial(\mathbf{e}_{TE} \cdot \mathbf{e}^*_{TM})}{\partial x} \right] dxdy. \quad (S14)$$

A similar procedure can be applied to evaluate the optical torque exerted on waveguide surfaces. Consider air-waveguide interface at $x = x_0$ plane, the derivative of normal electric field component ($E_x$) with respect to $x$ is a delta function, which leads to the surface force density

$$\mathbf{f}^{d,surf} = (\mathbf{P} \cdot \nabla)\mathbf{E} = \frac{1}{2}\varepsilon_0(\varepsilon-1)E_{x,mat}\left(E_x(x_0^+) - E_x(x_0^-)\right)\delta(x-x_0)\hat{\mathbf{x}}, \quad (S15)$$

where $E_{x,mat}$ is the $x$ component of the electric field inside the waveguide and $\hat{\mathbf{x}}$ is the unit vector towards $x$ axis. Equation (S15) describes a surface force density with its direction normal to the interface. With the boundary condition on the normal component of $\mathbf{E}$, we have

$$\varepsilon E_{x,mat} = E_{x,air}. \quad (S16)$$

Therefore, Eq. (S15) can be rewritten as

$$\mathbf{f}^{d,surf} = (\mathbf{n} \cdot \hat{\mathbf{x}})\frac{1}{2}\varepsilon_0(\varepsilon-1)^2 E_{x,mat}^2 \delta(x-x_0)\hat{\mathbf{x}}, \quad (S17)$$

where **n** is the unit vector towards outside waveguide. The associated torque exerted on the surface $x_0$ is

$$T^{d,surf}(x_0) = -\int y f^{d,surf} d^3x = -(\mathbf{n} \cdot \hat{\mathbf{x}}) \frac{1}{2} \varepsilon_0 (\varepsilon - 1)^2 \int y E_{x,mat}^2(x_0) dy dz . \tag{S18}$$

Using complex field expression, the time-averaged torque on surface $x_0$ is

$$\langle T^{d,surf}(x_0) \rangle = \int dz \left\{ -(\mathbf{n} \cdot \hat{\mathbf{x}}) \frac{1}{2} \varepsilon_0 (\varepsilon - 1)^2 \cos(\varphi) \int y E_{x,mat,TE}(x_0) E_{x,mat,TM}^*(x_0) dy \right\} = \int dz \langle \tau^{d,surf}(z) \rangle , \tag{S19}$$

where $\langle \tau^{d,surf}(z) \rangle$ can be written in terms of the normalized field amplitudes as

$$\langle \tau^{d,surf}(z) \rangle = -(\mathbf{n} \cdot \hat{\mathbf{x}}) \frac{1}{2} \varepsilon_0 (\varepsilon - 1)^2 \cos(\varphi)(a_x a_y) \int y e_{x,mat,TE}(x_0) e_{x,mat,TM}^*(x_0) dy . \tag{S20}$$

The result Eq. (S20) illustrates that optical torque on waveguide surfaces shows the same sinusoidal dependence on $\varphi$ as its bulk counterpart in Eq. (S12). In comparison with Eq. (S12) and Eq. (S14), one can easily obtain surface linear torque density $\langle \tau^{d,surf}(z) \rangle$ and the corresponding coefficient $\eta^{d,surf}$. Therefore, the total linear torque density that arises from dipole force simply reads

$$\langle \tau^d(z) \rangle = \langle \tau^{d,bulk}(z) \rangle + \langle \tau^{d,surf}(z) \rangle . \tag{S21}$$

**1.b. Electrostriction**

For isotropic materials and for cubic crystals (such as silicon), the electrostrictively-induced stress is related to the optical fields as[2]

$$\langle \sigma_{kl}^{es} \rangle = -\frac{1}{4} \varepsilon_0 n^4 p_{ijkl} \operatorname{Re}(E_i E_j^*) , \tag{S22}$$

and the force density is calculated from the stress tensor as

$$\langle f_i^{es} \rangle = -\partial_j \langle \sigma_{ji}^{es} \rangle . \tag{S23}$$

Due to the discontinuity of $\langle \sigma_{ji}^{es} \rangle$ at the material boundaries, $\langle f_i^{es} \rangle$ needs to be evaluated inside the material and at the dielectric interfaces separately, similar to the dipole force case. All of the stress components that are related to torque calculation are listed below:

$$\begin{aligned}\langle \sigma_{xx}^{es} \rangle &= -\frac{1}{4} \varepsilon_0 n^4 (p_{11} |E_x|^2 + p_{21} |E_y|^2 + p_{31} |E_z|^2) \\ &= -\frac{1}{4} \varepsilon_0 n^4 (p_{11} E_{x,TE} E_{x,TM} + p_{21} E_{y,TE} E_{y,TM} + p_{31} E_{z,TE} E_{z,TM}) e^{i\varphi} + c.c.\end{aligned} \tag{S24}$$

$$\langle \sigma_{xy}^{es} \rangle = \langle \sigma_{yx}^{es} \rangle = -\frac{1}{4}\varepsilon_0 n^4 p_{66}(E_x E_y^* + E_y E_x^*)$$
$$= -\frac{1}{4}\varepsilon_0 n^4 p_{66}(E_{x,\text{TE}} E_{x,\text{TM}} + E_{x,\text{TM}} E_{y,\text{TE}})e^{i\varphi} + c.c. \quad (S25)$$

$$\langle \sigma_{zx}^{es} \rangle = -\frac{1}{4}\varepsilon_0 n^4 p_{55}(E_z E_x^* + E_x E_z^*)$$
$$= -\frac{1}{4}\varepsilon_0 n^4 p_{55}(iE_{z,\text{TE}} E_{x,\text{TM}} - iE_{z,\text{TM}} E_{x,\text{TE}})e^{i\varphi} + c.c. \quad (S26)$$

$$\langle \sigma_{yy}^{es} \rangle = -\frac{1}{4}\varepsilon_0 n^4 (p_{12}|E_x|^2 + p_{22}|E_y|^2 + p_{32}|E_z|^2)$$
$$= -\frac{1}{4}\varepsilon_0 n^4 (p_{12} E_{x,\text{TE}} E_{x,\text{TM}} + p_{22} E_{y,\text{TE}} E_{y,\text{TM}} + p_{32} E_{z,\text{TE}} E_{z,\text{TM}})e^{i\varphi} + c.c. \quad (S27)$$

$$\langle \sigma_{zy}^{es} \rangle = -\frac{1}{4}\varepsilon_0 n^4 p_{44}(E_y E_z^* + E_y^* E_z)$$
$$= -\frac{1}{4}\varepsilon_0 n^4 p_{44}(-iE_{y,\text{TE}} E_{z,\text{TM}} + iE_{y,\text{TM}} E_{z,\text{TE}})e^{i\varphi} + c.c. \quad (S28)$$

Therefore, the corresponding electrostrictive force and torque density inside the silicon waveguide are:

$$\langle f_x^{es,bulk} \rangle = -\frac{\partial \langle \sigma_{xx}^{es} \rangle}{\partial x} - \frac{\partial \langle \sigma_{yx}^{es} \rangle}{\partial y} - \frac{\partial \langle \sigma_{zx}^{es} \rangle}{\partial z}$$
$$= \frac{1}{2}\varepsilon_0 n^4 \left[ \frac{\partial}{\partial x}\left( p_{11} E_{x,\text{TE}} E_{x,\text{TM}} + p_{21} E_{y,\text{TE}} E_{y,\text{TM}} + p_{31} E_{z,\text{TE}} E_{z,\text{TM}} \right) \right.$$
$$+ \frac{\partial}{\partial y}\left( p_{66} E_{x,\text{TE}} E_{y,\text{TM}} + p_{66} E_{x,\text{TM}} E_{y,\text{TE}} \right)$$
$$\left. + \Delta k \left( -p_{55} E_{z,\text{TE}} E_{x,\text{TM}} + p_{55} E_{z,\text{TM}} E_{x,\text{TE}} \right) \right] \cos(\varphi) \quad (S29)$$

$$\langle f_y^{es,bulk} \rangle = -\frac{\partial \langle \sigma_{xy}^{es} \rangle}{\partial x} - \frac{\partial \langle \sigma_{yy}^{es} \rangle}{\partial y} - \frac{\partial \langle \sigma_{zy}^{es} \rangle}{\partial z}$$
$$= \frac{1}{2}\varepsilon_0 n^4 \left[ \frac{\partial}{\partial x}\left( p_{66} E_{x,\text{TE}} E_{y,\text{TM}} + p_{66} E_{x,\text{TM}} E_{y,\text{TE}} \right) \right.$$
$$+ \frac{\partial}{\partial y}\left( p_{12} E_{x,\text{TE}} E_{x,\text{TM}} + p_{22} E_{y,\text{TE}} E_{y,\text{TM}} + p_{32} E_{z,\text{TE}} E_{z,\text{TM}} \right)$$
$$\left. + \Delta k \left( p_{44} E_{y,\text{TE}} E_{z,\text{TM}} - p_{44} E_{y,\text{TM}} E_{z,\text{TE}} \right) \right] \cos(\varphi) \quad (S30)$$

$$\langle \tau^{es,bulk}(\mathbf{r}) \rangle = x \langle f_y^{es,bulk} \rangle - y \langle f_x^{es,bulk} \rangle \quad (S31)$$

The corresponding electrostrictive force and torque density on the silicon waveguide surface can be expressed in a similar way.

The main feature of electrostrictively-induced linear torque density is the same sinusoidal dependence on $\varphi$, as shown in Eq. (S29), (S30) and (S31). Therefore, a similar set of quantities such as $S_e^{es}$ and $\eta^{es}$ can be derived from them.

By summing up the optical torque from dipole and electrostrictive contributions, we obtain Eq. (1) and Eq. (2) in the main text for total linear torque density and the effective spin angular momentum of photon in media.

### 1.c. Numerical calculations

Having derived the expression for optical torque density resulted from two different effects, we proceed to calculate the optical torque in a suspended silicon waveguide. We consider a silicon waveguide with 400 nm in width and 340 nm in height, surrounded by air. Figure S1b, c and d illustrate the torque density profiles inside the waveguide when the two modes are in phase ($\varphi = 0$) and equal in power ($a_x = a_y$).

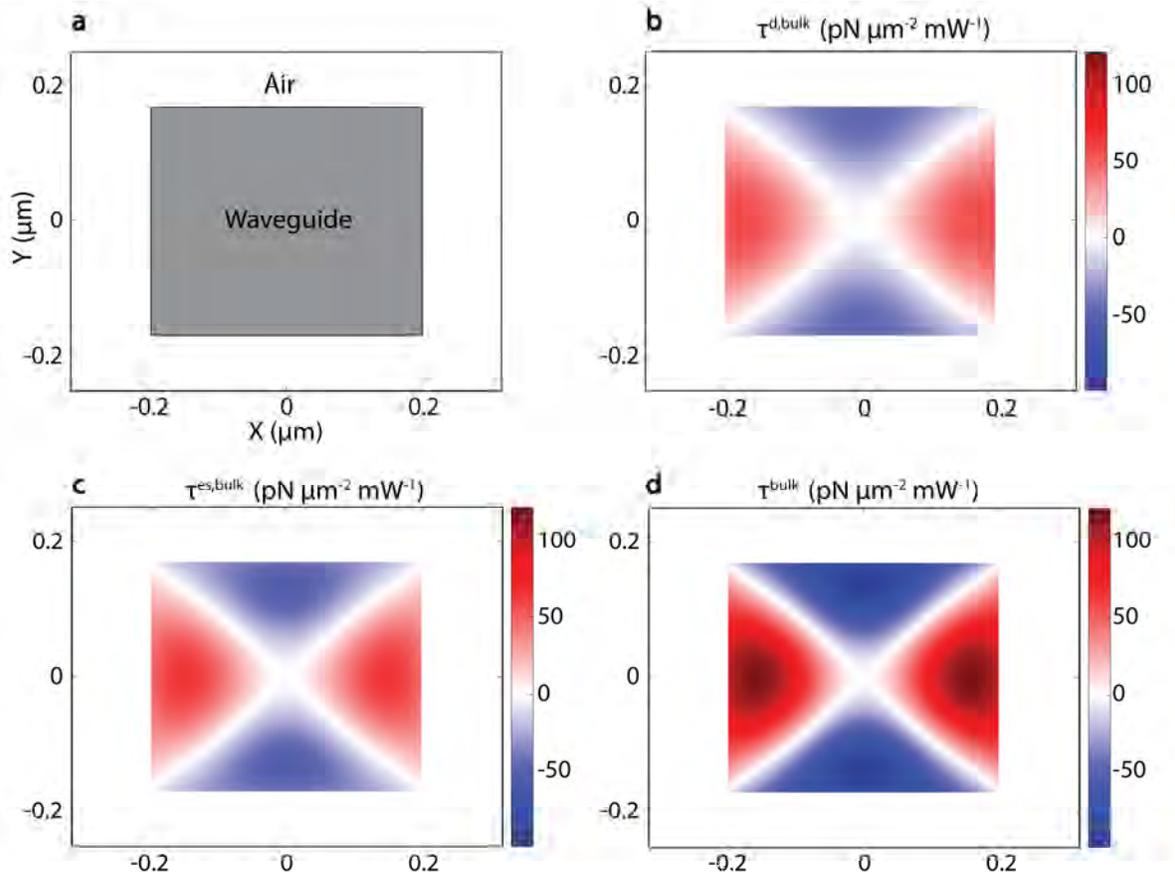

**Figure S1. Simulated torque density distributions inside a silicon waveguide (bulk contribution) suspended in air. a**. The waveguide geometry is 400 nm wide and 340 nm high. **b**. **c**. **d**. Torque density distributions due to **b**. dipole force, **c**. electrostrictive force and **d**. the sum of the two.

Figure S2 depicts the non-vanishing surface torque density on the air-waveguide interface for both dipole and electrostrictive effects due to the discontinuity of normal electric field and stress tensor across the boundaries.

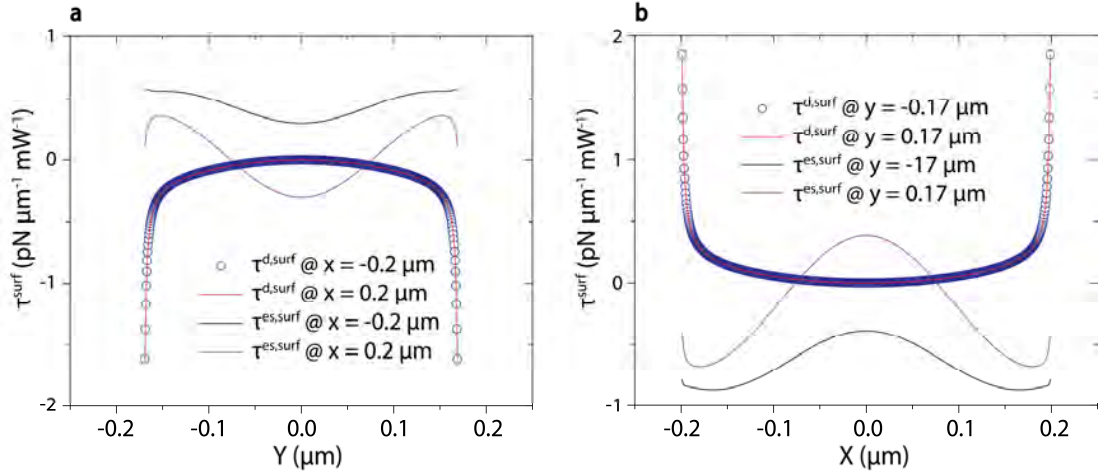

**Figure S2. Simulated surface torque density distribution along waveguide surfaces.** The dipole and electrostrictive contributions are plotted separately.

To explore the dependence of optical torque on waveguide geometry, we calculate the coefficient $\eta$ for waveguides with various widths. As shown in Fig. S3, the coefficient $\eta$, or effective spin angular momentum $S_e(\varphi=0)$, varies slightly as the waveguide width increases. In addition, the dipole ($\eta^d$) and electrostrictive contributions ($\eta^{es}$) to the optical torque are of similar magnitude.

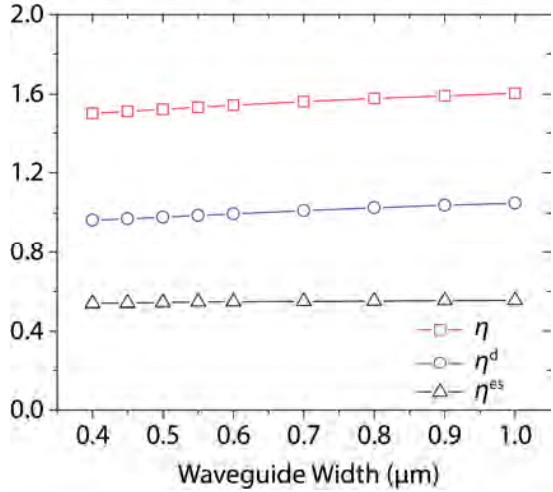

**Figure S3. Simulated coefficient $\eta$ for waveguides with various width, while waveguide height is fixed to be 340 nm.**

## 2. Photonic Crystal Nanocavity Design

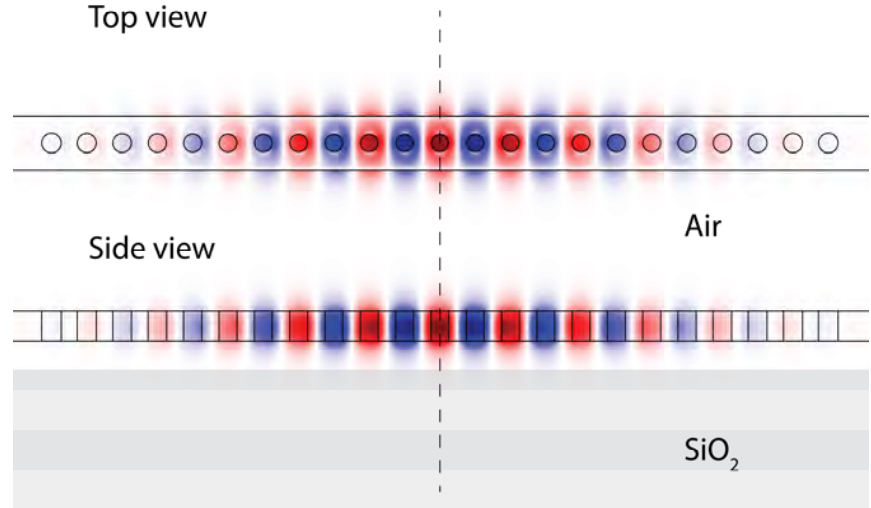

**Figure S4. Simulated mode profile (transverse electric field) of the photonic crystal nanocavity.**

The two photonic crystal nanocavities on the suspended nanobeam are designed to be identical using a deterministic method[3]. In order to achieve the optimal trade-off between high quality factors and strong optomechanical coupling, the fundamental air mode is used. Each nanocavity is 600 nm wide, 340 nm thick and 27.6 µm long on one side of the suspended waveguide, as shown in Fig. 2c. A total of 69 holes are used in each nanocavity and the lattice constant (the distance between the centers of adjacent holes) is 400 nm. The diameter of the holes varies from 240 nm, at the edges of the cavity, to 172 nm, at the center of the cavity. The simulated mode profile is shown in Fig. S4.

## 3. Waveguide-Nanobeam Junction Design

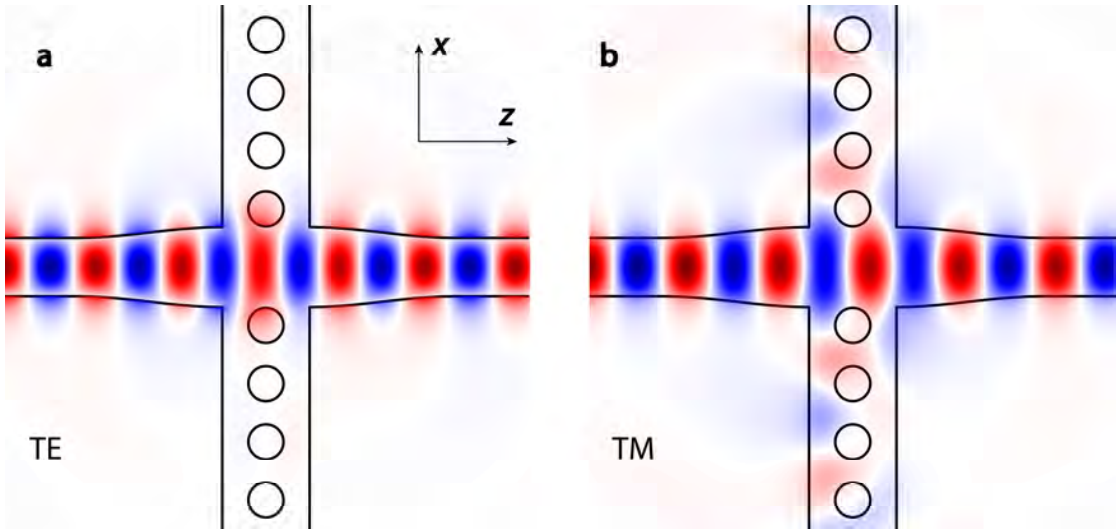

**Figure S5. FDTD simulation results showing the transmission of TE and TM mode through the junction structure.** We plot $E_x$ and $E_y$ component for TE and TM mode, respectively.

The junction formed by the waveguide and photonic crystal nanobeams is designed to minimize optical reflection or scattering for both TE and TM modes due to the abrupt change in structure geometry. To reduce the mode mismatch, the waveguide width is first expanded from 400 nm to 550 nm on the input side before joining with the nanobeams and tapered back to 400 nm symmetrically on the output side. The position of the photonic crystals with respect to the wavegudie central axis is fine tuned using finite-difference time-domain (FDTD) calculaitons to avoid pump light leakage from the waveguide to photonic crystal cavities. In Fig. S5 we plot the simulated mode profiles when TE and TM mode are propagating through the junction structure. For TE mode, since pump light wavelength is within the bandgap of the photonic crystals aside, light leakage from the waveguide into the two photonic crystal cavities is forbidden. On the other hand, there is a small amount of scattering light for TM mode due to the lack of bandgap confinement. However, the weak leakage of light from the TM mode should not affect our measurement, as pump light wavelength (1540 nm) is far away from the photonic crystal cavity resonance (~1530 nm).

### 4. TE to TM Mode Converter Design

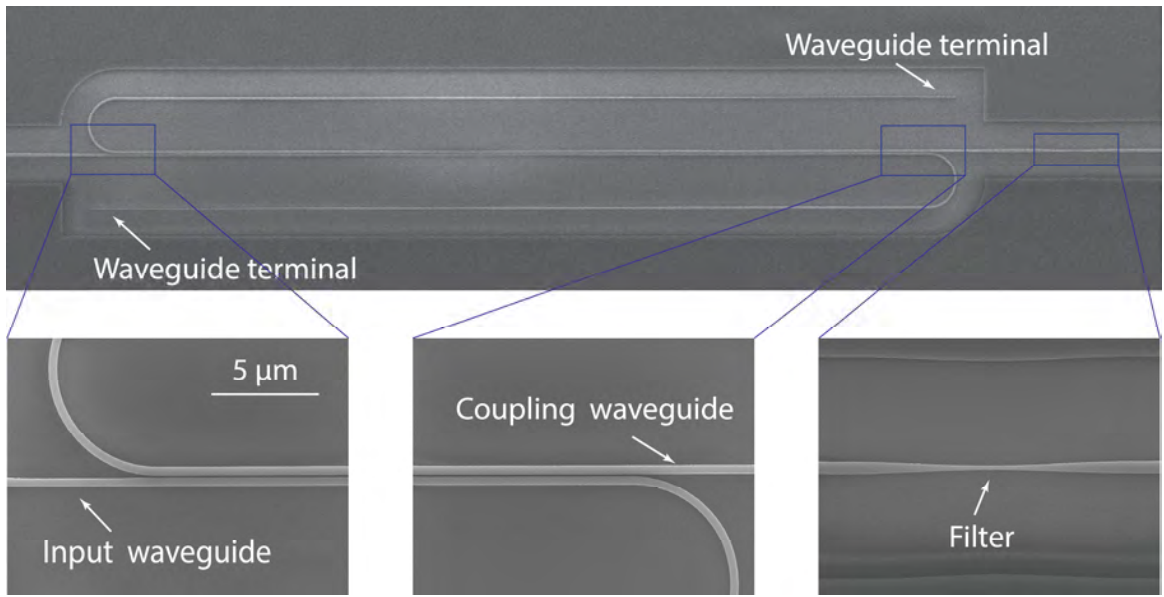

**Figure S6. Scanning electron micrograph of the mode convertor.**

The mode conversion from TE to TM mode is achieved with an asymmetric directional coupler[4] shown in Fig. S6. The width of the input and coupling waveguides are 410 nm and 430 nm, respectively, and the gap of the two is about 100 nm. In order to satisfy the phase matching condiiton, the effective index of the fundamental TE mode for the input waveguide is desigend to close to it of the fundamental TM mode for the coupling waveguide. During the mode progation in the two coupled waveguides, the fundametnal TE mode in the input waveguide is evanecently coupled into the fundamental TM mode of the coupling waveguide. To filter out any possible higher order modes excited in the coupling waveguide other than the fundamental TM mode, the coupling waveguide after the mode convertor is first narrowed down to 200 nm wide so that only fundamental TM mode is allowed to propagte. The wavaeguide is then expanded back symmetrically and join with the path split from the TM input GC to form the interferometer. The input waveguide is adiabatically tapered down to 80 nm to terminate all residual light without reflection.

## 5. Simulation of mechanical modes

We used Finite Element Method (FEM) to simulate the mechanical modes supported by the suspended silicon waveguide and nanobeam, taking into account the anisotropic elasticity of silicon[5]. Our devices were fabricated from a {100} SOI wafer, such that the wafer surface normal ($y$ axis) is along the $\langle 100 \rangle$ direction, while the suspended waveguide ($z$ axis) and nanobeam ($x$ axis) are both along $\langle 110 \rangle$ directions. (The coordinates are defined in Fig. 3b.) The simulated profiles of the four experimentally observed mechanical modes in Fig. 3c are shown in Fig. S7.

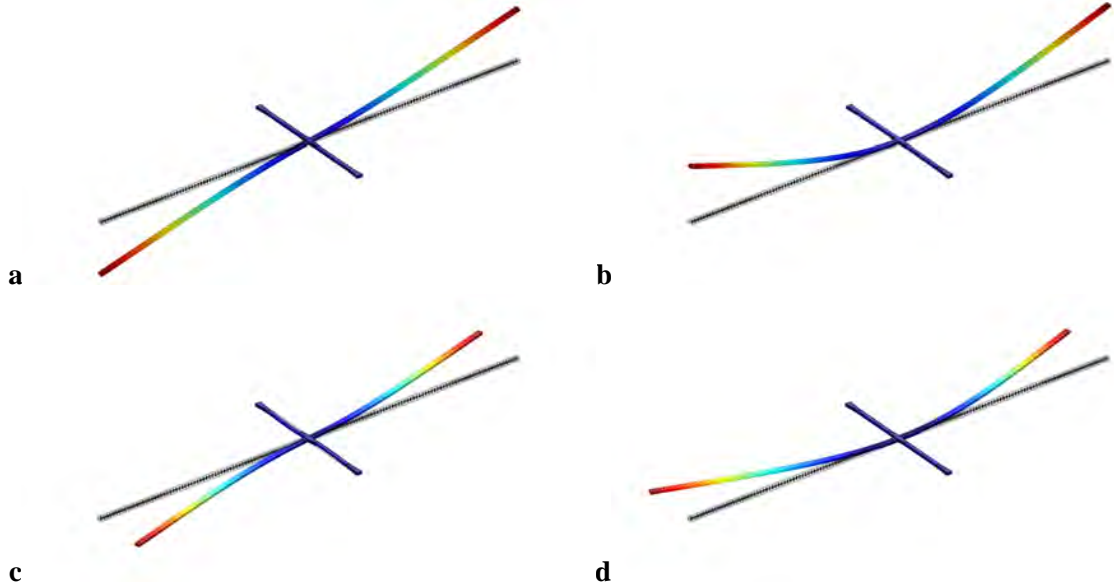

a

b

c

d

**Figure S7. Simulated mechanical mode profiles of the suspended silicon waveguide and nanobeam.** **a**. Fundamental (first order) out-of-plane torsional (anti-symmetric) mode. **b**. Fundamental (first order) out-of-plane flapping (symmetric) mode. **c**. Fundamental (first order) in-plane torsional (anti-symmetric) mode. **d**. Fundamental (first order) in-plane flapping (symmetric) mode.

In Fig. 3e, the measurement results show ~10% lower frequency than simulation using typical value of bulk silicon's elasticity matrix, indicating the elasticity matrix of the silicon layer in the SOI is effectively lower. Therefore, we fitted the measurement results by reducing every element in the elasticity matrix by 19% and achieved excellent agreement. The simulation results obtained with this reduced elasticity matrix were used for the subsequent analysis. (It is worth noting that scaling every element of the elasticity matrix by the same factor does not change the normalized mode profiles.)

## 6. Effective torsional simple harmonic oscillator model

In our experiment, we focused only on the fundamental torsional mode shown in Fig. 3b, where the waveguide undergoes pure torsional motion with its two ends fixed, while the nanobeam undergoes not only pure rotation with respect to $z$ axis, but also non-negligible deformation from a straight nanobeam, due to its finite rigidity. In addition to the full time-dependent 3D vector field description $\mathbf{u}(\mathbf{r},t)$ of this mode, it can be fully and accurately described by two time-dependent scalar functions, $\theta(z,t)$ and $v(x,t)$, which are the rotation angle of the waveguide as a function of $z$ and the vertical displacement of the nanobeam as a function of $x$, respectively, because both the waveguide and the nanobeam are very thin and long. Due to the continuity of the elastic material,

$$\left.\theta(z,t)\right|_{z=0} = \frac{d}{dx} v(x,t) \bigg|_{x=0}, \text{ when } \left.\theta(z,t)\right|_{z=0} = \theta(0,t) \ll 1. \tag{S32}$$

The mode profile deviates from that of an ideal torsional simple harmonic oscillator (SHO), which should possess massless torsional spring and rigid torsional mass. Therefore, in order to accurately quantify the excitation and response of this mode, we employ an effective torsional SHO model[6], whose equation of motion is

$$I_e \ddot{\Theta}(t) + \frac{\sqrt{I_e \kappa_e}}{Q} \dot{\Theta}(t) + \kappa_e \Theta(t) = T_e(t), \tag{S33}$$

where $\Theta$ is the instantaneous mode amplitude in terms of a rotation angle, the overhead dots denote time derivatives, $I_e$ is the effective moment of inertia, $\kappa_e$ is the effective torsional spring constant, $Q$ is the quality factor and $T_e$ is the effective torque. Although theoretically $\Theta$ can be arbitrarily defined, we choose to define it as the rotation angle at the center of the nanobeam, that is

$$\Theta(t) = \theta(0,t). \tag{S34}$$

Following this definition, the normalized mode profiles $\mathbf{u}_n(\mathbf{r})$, $\theta_n(z)$ and $v_n(x)$ can be defined as

$$\mathbf{u}(\mathbf{r},t) = \Theta(t) \mathbf{u}_n(\mathbf{r}), \tag{S35}$$

$$\theta(z,t) = \Theta(t) \theta_n(z), \text{ which requires that } \theta_n(0) = 1, \text{ and} \tag{S36}$$

$$v(x,t) = \Theta(t) v_n(x), \text{ which requires that } \frac{d}{dx} v_n(x) \bigg|_{x=0} = 1. \tag{S37}$$

The simulated $\theta_n(z)$ and $v_n(x)$ are summarized in Fig. S8. Furthermore, $I_e$ and $\kappa_e$ can be calculated using

$$U_k(t) = \frac{1}{2} I_e \dot{\Theta}^2(t) \text{ and } U_p(t) = \frac{1}{2} \kappa_e \Theta^2(t), \tag{S38}$$

where $U_k(t)$ and $U_p(t)$ are, respectively, the simulated instantaneous kinetic and potential energy corresponding to $\Theta(t)$. Naturally, the simulated mode (angular) frequency $\Omega_0$ satisfies

$$\Omega_0 = 2\pi f_0 = \sqrt{\kappa_e / I_e}. \tag{S39}$$

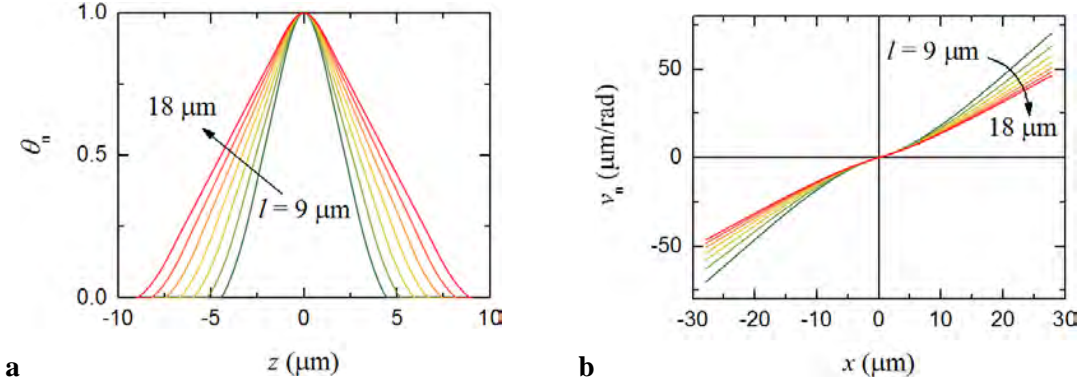

**Figure S8. Simulated normalized mode profiles of the suspended silicon waveguide and nanobeam.** The total waveguide length $l$ is varied from 9 μm to 18 μm, in steps of 1.5 μm, consistent with the actual devices. **a**. Normalized rotation angle of the waveguide as a function of $z$. **b**. Normalized vertical displacement of the nanobeam as a function of $x$.

The simulated parameters of the effective torsional SHO are summarized in Table S1.

**Table S1. Simulated parameters of the effective torsional SHO.**
Abbreviations: Symm, symmetric; Anti-symm, anti-symmetric.

| $l$ (μm) | 9.0 | 10.5 | 12 | 13.5 | 15 | 16.5 | 18.0 |
|---|---|---|---|---|---|---|---|
| $I_e$ (ng·μm$^2$) | 30.63 | 24.84 | 21.09 | 18.51 | 16.65 | 15.25 | 14.16 |
| $\kappa_e$ (pN·μm/μrad) | 173.3 | 128.9 | 101.0 | 82.19 | 68.95 | 59.17 | 51.66 |
| $f_0$ (kHz) | 378.6 | 362.6 | 348.2 | 335.3 | 323.9 | 313.5 | 304.0 |

## 7. Calibration of optomechanical measurement transduction factors

We used the measured thermo-mechanical noise (TMN) power spectral density (PSD)[6,7] to determine the resonance frequency $f_0$ and the quality factor $Q$ of the effective torsional SHO, and most importantly, to calibrate the transduction of the torsional motion.

When the pump laser is absent and the probe laser power is sufficiently attenuated to induce negligible dynamic cavity optomechanical backaction, the effective torsional SHO of the suspended structure is thermally excited at room temperature ($T_{lab}$ = 300 K), such that the expectation value of its kinetic and potential energy

$$\langle U_k(t) \rangle = \langle U_p(t) \rangle = \frac{1}{2} k_B T_{lab}, \tag{S40}$$

where $k_B$ is the Boltzmann constant. Therefore, the effective torsional SHO undergoes Brownian motion, which, according to Eq. (S38), can be quantified as

$$\langle \Theta^2(t) \rangle = \frac{k_B T_{lab}}{\kappa_e}. \tag{S41}$$

Assuming that $Q$ is high, the squared modulus of the transfer function of the effective torsional SHO is well approximated as a Lorentzian function. Further assuming that the PSD of the thermal excitation is white, i.e., constant in the frequency range of interest, the PSD of the Brownian motion of the effective torsional SHO is also a Lorentzian function, the integration of which should be equal to Eq. (S41).

The torsional motion $\Theta(t)$ is transduced by the probe laser and the photodetector into a voltage signal $V(t)$ and measured by a spectrum analyzer (SA). We define the transduction factor $G$ such that

$$V(t) = V_{\text{Bias}} + G\Theta(t), \tag{S42}$$

where the DC bias voltage is filtered out and the SA only measures the $G\Theta(t)$ term. The transduction factor $G$ thus defined is determined by the optomechanical coupling of the nanocavity, the optical resonance line shape, the grating coupler efficiency, the probe laser power and the photodetector gain. In the actual measurement, we configured the SA to display the PSD in terms of $V/\sqrt{\text{Hz}}$ versus single-sided frequency $f$, such that the raw data measured by the SA and plotted in Fig. 3c is $\sqrt{S_{VV}^{\text{SA}}(f)}$. For the mode of interest,

$$S_{VV}^{\text{SA}}(f) = G^2 \left[ \frac{\frac{k_B T_{\text{lab}}}{4\pi^2 \sqrt{I_e \kappa_e} Q}}{(f - f_0)^2 + \left(\frac{f_0}{2Q}\right)^2} \right], \tag{S43}$$

which is used for the curve fitting to obtain $f_0$, $Q$ and $G$. The calibrated $G$ is used to converted the measured voltage into $\Theta(t)$.

## 8. Effective torque

When an arbitrary time-dependent external force volume density distribution $\mathbf{f}(\mathbf{r},t)$ is applied on the structure, effective torque $T_e(t)$ should be calculated as the overlap integral[6,7] between $\mathbf{f}(\mathbf{r},t)$ and $\mathbf{u}_n(\mathbf{r})$. In our case, $\mathbf{f}(\mathbf{r},t)$ is exerted by the guided optical modes inside the waveguide, hence is negligible inside the nanobeam. Therefore,

$$T_e(t) = \iiint_{W \cup B} \mathbf{f}(\mathbf{r},t) \cdot \mathbf{u}_n(\mathbf{r}) dV \approx \iiint_W \mathbf{f}(\mathbf{r},t) \cdot \mathbf{u}_n(\mathbf{r}) dV, \tag{S44}$$

where $W$ and $B$ represent the regions of the waveguide and the nanobeam, respectively. Because the waveguide undergoes pure torsional motion,

$$\mathbf{u}_n(\mathbf{r}) \approx (\mathbf{z} \times \mathbf{r})\theta_n(z) \text{ inside the waveguide, and} \tag{S45}$$

$$T_e(t) \approx \iiint_W \mathbf{f}(\mathbf{r},t) \cdot (\mathbf{z} \times \mathbf{r})\theta_n(z) dV = \iiint_W t_z(\mathbf{r},t)\theta_n(z) dV, \tag{S46}$$

which can be further simplified as

$$T_e(t) \approx \int_W \left[ \iint_W \mathbf{t}_z(\mathbf{r},t) dxdy \right] \theta_n(z) dz \approx \int_W \tau(z,t) \theta_n(z) dz . \tag{S47}$$

In order to derive a closed-form expression for $T_e(t)$, we approximate the waveguide as a perfect rectangular waveguide with a constant cross-sectional size. Therefore, substituting Eq. (1) in the main text for $\tau(z,t)$ in Eq. (S47) yields

$$T_e(t) \approx \eta(\Delta k/\omega)(2a_x a_y) \int_{-l/2}^{l/2} \cos(\varphi(z,t)) \theta_n(z) dz, \text{ where } \varphi(z,t) = \varphi_0(t) + \Delta k z \tag{S48}$$

and $l$ is the total waveguide length, as defined in the main text. In our experiment we sinusoidally modulated $\varphi_0$, so we explicitly express its time-dependence in Eq. (S48) and the following derivations. The integrand in Eq. (S48) can be expanded as

$$\cos(\varphi_0(t))\cos(\Delta k z)\theta_n(z) - \sin(\varphi_0(t))\sin(\Delta k z)\theta_n(z),$$

where the first term is an even function of $z$, while the second term is an odd function of $z$, hence has no contribution to the integral, because $\theta_n(z)$ is an even function. Consequently the expression for $T_e(t)$ can be further simplified as

$$T_e(t) \approx T_m \cos(\varphi_0(t)), \text{ where} \tag{S49}$$

$$T_m = \eta(\Delta k/\omega)(2a_x a_y) \left[ 2\int_0^{l/2} \cos(\Delta k z)\theta_n(z) dz \right]. \tag{S50}$$

To proceed, we approximate $\theta_n(z)$ as a triangular function, namely

$$\theta_n(z) \approx 1 - |2z/l|, \text{ when } |z| \le l/2, \text{ and } \theta_n(z) = 0 \text{ otherwise.} \tag{S51}$$

By substituting Eq. (S51) into Eq. (S50), we solve the integral by parts and obtain the final closed-form expression for $T_m$

$$T_m \approx \eta \frac{(2a_x a_y)}{\omega} \left[ \frac{2\sin^2(\Delta kl/4)}{\Delta kl/4} \right] = -S_e\left(\frac{\pi}{2}\right) \Phi_q \frac{2\sin^2(\Delta kl/4)}{\Delta kl/4}, \tag{S52}$$

where $S_e$ and $\Phi_q$ are the effective photon spin angular momentum and the photon flux defined in the main text.

### 9. Resonance response of the effective torsional SHO

The resonance response of the effective torsional SHO can be derived from the transfer function of the effective torsional SHO, which, according to Eq. (S33), is

$$H(\Omega) = \frac{1}{-\Omega^2 I_e + j\Omega C + \kappa_e}, \text{ where } C = \frac{\sqrt{I_e \kappa_e}}{Q}. \tag{S53}$$

(In this section we adopt the conventions and terminology generally used in the field of electrical engineering, consistent with the network analyzer used in our measurement.) When $Q$ is large, $H(\Omega)$ can be very well approximated as

$$H(\Omega) = \frac{-\frac{1}{2CQ}}{(\Omega - \Omega_0) - j\frac{\Omega_0}{2Q}}. \tag{S54}$$

The real and imaginary parts of $H(\Omega)$ are,

$$\text{Re}(H(\Omega)) = -\frac{\frac{1}{2CQ}(\Omega - \Omega_0)}{(\Omega - \Omega_0)^2 + \left(\frac{\Omega_0}{2Q}\right)^2} \text{ and } \text{Im}(H(\Omega)) = -\frac{\frac{\Omega_0}{4CQ^2}}{(\Omega - \Omega_0)^2 + \left(\frac{\Omega_0}{2Q}\right)^2}, \tag{S55}$$

which correspond to the in-phase and quadrature components of the resonance response. As stated in the main text, for a sinusoidally modulated $\varphi_0$ that is $\delta\varphi_0 \cos(\Omega t)$, the corresponding sinusoidally modulated $T_e$ is $-T_m \sin(\varphi_0)\delta\varphi_0 \cos(\Omega t)$, therefore the measured quadrature component of the resonance response is

$$\Theta_{qd}(\Omega) = \left(-T_m \sin(\varphi_0)\delta\varphi_0\right)\text{Im}(H(\Omega)), \tag{S56}$$

which can be written in terms of $f$, instead of $\Omega$, to fit the data shown in Fig. 4a as

$$\Theta_{qd}(f) = \frac{\frac{f_0}{8\pi Q\sqrt{I_e \kappa_e}}\left(T_m \sin(\varphi_0)\delta\varphi_0\right)}{(f - f_0)^2 + \left(\frac{f_0}{2Q}\right)^2}, \tag{S57}$$

where the term $T_m \sin(\varphi_0)$ can be obtained for each curve. Clearly, the sign of $\Theta_{qd}(f)$ depends solely on the sign of $\sin(\varphi_0)$.